\documentclass{elsart}
\usepackage{amsfonts}
\usepackage{amsmath}
\usepackage{graphicx}
\usepackage{amssymb}

\def\comment#1{}

\def\mn#1{*\marginpar{*\tiny{#1}}}
\def\mn#1{}
\def\E{{\mathcal E}}

\begin{document}
\begin{frontmatter}
\title{Electron-positron pair oscillation in spatially inhomogeneous electric fields
and radiation}
\author
{Wen-Biao Han, Remo Ruffini, She-Sheng Xue}
\address
{ICRANet Piazzale della Repubblica, 10-65122, Pescara, \\and Physics Department, University of Rome "La Sapienza," P.le A. Moro 5, 00185 Rome, Italy}
\comment{\author{W.-B.~Han, H.~Kleinert,
R.~Ruffini and S.-S. Xue}
\address{ICRANet, P.le della Repubblica 10, 65100 Pescara, Italy, \\
ICRA and University of Rome "Sapienza", Physics Department, \\
P.le A. Moro 5, 00185 Rome, Italy.}}
\begin{abstract}
It is known that strong electric fields produce electron and positron pairs from the vacuum, and due to the back-reaction these pairs oscillate back and forth coherently with the alternating electric fields in time.
We study this phenomenon in spatially inhomogeneous and bound electric fields by
integrating the equations of energy-momentum and particle-number
conservations and Maxwell equations. The space and time evolutions
of the pair-induced electric field, electric charge- and
current-densities are calculated. The results show that
non-vanishing electric charge-density and the propagation of
pair-induced electric fields, differently from the case of
homogeneous and unbound electric fields. The space and time
variations of pair-induced electric charges and currents emit an
electromagnetic radiation. We obtain the narrow spectrum and
intensity of this radiation, whose peak $\omega_{\rm peak}$ locates
in the region around $4$ KeV for electric field strength $\sim E_c$.
We discuss their relevances to both the laboratory experiments for
electron and positron pair-productions and the astrophysical
observations of compact stars with an electromagnetic structure.
\comment{ Based on the Maxwell equations, energy-momentum and
particle-number conservations, we study the back-reaction of
electron-positron pairs created in spatially inhomogeneous electric
fields.  Numerically integrating a set of partial differential
equations, we find, in inhomogeneous case, there would be net
charges at local slices instead of everywhere is neutral in
spatially uniform electric field. Then the involved physical
variables (electric field, number density, energy density and
momentum density of positrons and electrons etc.)can evolve to more
complicated sub structure in space. The most interesting thing is
that the electric field and current not only synchronously oscillate
in time, but also there is a wave propagating along the inverse
direction of external electric field gradient in space. }
\end{abstract}
\end{frontmatter}

\noindent
{\it Introduction.}
\hskip0.1cm
As reviewed in the recent report \cite{report}, since the pioneer
works by Sauter \cite{1}, Heisenberg and Euler \cite{2} in 1930's, then by
Schwinger \cite{3} in 1950's, it has been well known that positron-electron pairs are produced from the vacuum in external electric fields. In a constant electric field
$E_0$ in dependent of space and time, the pair-creation rate per unit volume is given by \cite{2,3},
\begin{equation}
S\equiv\frac{dN}{dVdt}=\frac{m_e^4}{4\pi^3}\left(\frac{E_0}{E_c}\right)^2\exp\left(-\pi\frac{E_c}{E_0}\right),
\label{srate}
\end{equation}
where the critical field $E_c\equiv m_e^2c^3/(e\hbar)$, the
Plank's constant $\hbar$, the speed of light $c$, the electron mass $m_e$,
the absolute value of electron charge $e$ and the fine structure constant $\alpha =e^2/\hbar c$ (in this article we use the natural units $\hbar=c=1$, unless otherwise specified).
The pair-production rate (\ref{srate}) is significantly large for
strong electric fields $E \gtrsim E_c\simeq  1.3\cdot 10^{16}{\rm
V}/{\rm cm}$. The critical field will probably be reached by recent
advanced laser technologies in laboratory experiments
\cite{Ringwald,Tajima,Gordienko}, X-ray free electron laser
(XFEL) facilities \cite{XFEL}, optical high-intensity laser
facilities such as Vulcan or ELI \cite{ELI}, and SLAC E144 using
nonlinear Compton scattering \cite{burke1997}. On the
other hand, strong overcritical electric fields ($E\geq 10E_c$) can be created in
astrophysical environments, for instance, quark stars
\cite{Usov1,Usov2} and neutron stars \cite{Xue4}-\cite{Xue3}.

The back-reaction and screening effects of electron and positron
pairs on external electric fields lead to the phenomenon of plasma
oscillations: electrons and positrons moving back and forth
coherently with alternating electric fields. This means that
external electric fields are not eliminated within the Compton time
$\hbar/m_ec^2$  of pair-production process, rather oscillate
collectively with the motion of pairs in a much longer timescale.

In a constant electric field $E_0$ (\ref{srate}), the phenomenon of
plasma oscillations is studied in the two frameworks \cite{report}: (1)
the semi-classical QED with quantized Dirac field and classical
electric field \cite{4,QFT}; (2) the kinetic description
using the Boltzmann-Vlasov and Maxwell equations
\cite{ion,Matsui1,ion2,5,gregory}.
In the second framework, the Boltzmann-Vlasov equation is used to obtain the equations for the continuity and energy-momentum conservations \cite{Matsui1}.

Ref.~\cite{gregory} shows the evidence of plasma oscillation in
under-critical field ($E<E_c$) and the relation between the kinetic
energy and numbers of oscillating pairs in a given electric field
strength $E_0$. Taking into account the creation and annihilation
process $e^++e^-\Leftrightarrow \gamma+\gamma$, it is shown \cite{5}
that the plasma oscillation in an overcritical field is led to a
plasma of photons, electrons and positions with the equipartition of
their number- and energy-densities. The phenomenon of plasma
oscillations is studied in connection with pair creation in heavy ions
collisions \cite{ion}-\cite{ion2}, the laser field
\cite{Laser}-\cite{Laser3}, and gravitational collapse \cite{Luca}.
It is worthwhile to emphasize that the plasma oscillation occurs not only
at overcritical field-strengths $E_0\gtrsim E_c$ (see for instance Refs.~\cite{4,5}), but also undercritical field-strengths $E_0\lesssim E_c$ (see Ref.~\cite{gregory}), and plasma oscillation frequency
is related to field-strength $E_0$, while the number of oscillating pairs depends on the pair-production rate (\ref{srate}).
More details can be found in the recent review article \cite{report}.

The realistic ultra-strong electric fields are not only vary with
space and time, but also confined in a finite region. In
this letter, studying the plasma oscillations in
spatially inhomogeneous electric field, 
we present the evidence of electric
fields propagation, leading to electromagnetic radiation with a
peculiar narrow spectrum in the KeV-region, which should be distinctive
and experimentally observable.

\comment{
which is supposed to be
created by external sources, for instance, either the focusing spot of two
laser beams or astrophysical compact stars.
For space- or time-dependent electric fields, Kleinert et al
\cite{Hagen} calculated some formulas of pair production rates. So,
the results of Kleinert et al let us can study pair oscillation in
space-dependently inhomogeneous electric field. This is the main
object of this paper, and is very useful in practically physical and
astrophysical environment.
}

In the kinetic description for the plasma fluids of positrons ($+$) or electrons ($-$), whose single-particle spectrum $p_{\pm}^0=({\bf p}_\pm^2+m_e^2)^{1/2}$,
we define the number-densities $n_\pm (t,{\bf x})$ and ``averaged''
velocities  ${\bf v}_\pm (t,{\bf x})$ of the fluids:
\begin{align}
n_\pm (t,{\bf x}) &\equiv  \int  \frac{d^3{\bf p}_\pm}{(2\pi)^3}f_\pm(t,{\bf p}_\pm,{\bf x}),\label{meann}\\
{\bf v}_\pm (t,{\bf x}) &\equiv  \frac{1}{n_\pm}\int \frac{d^3{\bf p}_\pm}{(2\pi)^3} \left(\frac{{\bf p}_\pm} {p^0_\pm}\right) f_\pm(t,{\bf p}_\pm,{\bf x}),
\label{meanv}
\end{align}
where $f_\pm(t,{\bf p}_\pm,{\bf x})$ is the distribution function in the phase space. The four-velocities of the electron and positron fluids $U_\pm^{\mu}=\gamma_\pm (1,{\bf v}_\pm)$, the Lorentz factor $\gamma_{\pm}=( 1-|{\bf v}_{\pm}|^{2}) ^{-1/2}$, and the comoving number-densities $\bar n_\pm=n_\pm(\gamma_\pm)^{-1}$,
where we choose the laboratory frame where pairs are created at rest.
The collision-less plasma fluid of electrons and positrons coupling to electromagnetic fields is governed
by the continuity, energy-momentum conservation and Maxwell equations:
\setlength\arraycolsep{0.5pt}
\begin{align}
&\frac{\partial\left(  \bar{n}_\pm U_\pm^{\mu}\right)  }{\partial
x^{\mu}}
=S,\label{contp}\\
&\frac{\partial T_\pm^{\mu\nu}}{\partial x^{\nu}}
=-F^{\mu}_{\sigma}(J_\pm^{\sigma }+J_{\pm\rm pola}^{\sigma
}),\label{emp}\\
&\frac{\partial F^{\mu\nu}}{\partial x^{\nu}}   = -4\pi (J^{\mu}_{\rm cond}+J^{\mu}_{\rm pola}+J^{\mu}_{\rm ext}), \label{me}%
\end{align}
where $S=dN/dVdt$ is the pair-production rate, $J_\pm^\mu =\pm e\bar
n_\pm U^\mu_\pm$ electric currents and the energy-momentum tensors
\cite{Weinberg}
\begin{align}
T^{\mu\nu}_\pm &= \bar p_\pm g^{\mu\nu}+(\bar p_\pm +\bar \epsilon_\pm)U^\mu_\pm U^\nu_\pm,
\label{eptensor}
\end{align}
and the pressure $\bar p_\pm$ and comoving energy-density $\bar \epsilon_\pm$ is related by the equation of state, in general $0\le \bar p_\pm \le \bar \epsilon_\pm/3$. In the laboratory frame, the fluid energy-density
$\epsilon_{\pm} \equiv T^{00}$ and momentum-density $p^i_{\pm} \equiv  T^{i0}$ are given by
\begin{align}
\epsilon_{\pm} =(\bar\epsilon_{\pm}+\bar p_\pm {\bf v}^2_\pm)\gamma^2_{\pm},\quad
{\bf p}_{\pm} =(\bar\epsilon_{\pm}+\bar p_\pm )\gamma^2_{\pm}{\bf v}_\pm .
\label{led}
\end{align}
\comment{
\begin{align}
T_{\pm}^{\mu\nu}=m_e\bar{n}_{\pm}\left(  U_{\pm}^{\mu}U_{\pm}^{\nu}\right)  , \label{emten}%
\end{align}
given by the equation of motion $m_e\partial U_\pm^{\mu}/\partial z^\sigma = F^\mu_\sigma$.
}
In Eqs.~(\ref{emp},\ref{me}) $F^{\mu}_\sigma$ is the tensor of electromagnetic fields (${\bf E},{\bf B}$), the conducting four-current density
\begin{align}
J_{\rm cond}^{\mu}  &  \equiv e(\bar{n}_+U_+^{\mu} - \bar{n}_- U_-^{\mu}),\quad
\partial_\mu J_{\rm cond}^{\mu} =0,
\label{current}
\end{align}
and polarized four-current density $J_{\rm pola}^{\mu} = \sum_\pm
J_{\pm\rm pola}^{\mu}$ and $J_{\pm\rm pola}^{\mu}   =
\left(\rho^\pm_{\rm pola}, {\bf J}^\pm_{\rm pola} \right)$
\cite{Matsui1,Matsui2}
\begin{align}
F^\nu_\mu J_{\pm\rm pola}^{\mu}=\Sigma_\pm^\nu, \quad \Sigma^\nu_\pm \equiv
\int\frac{d^3{\bf p}_\pm}{(2\pi)^3p_\pm^0} p_\pm^\nu {\mathcal S},
\label{pcurrentd}
\end{align}
and $S=\int d^3{\bf p}_\pm/[(2\pi)^3p_\pm^0]{\mathcal S}$. Using ``averaged'' velocities (\ref{meanv}) of the fluids, we approximately have
\begin{align}
\quad {\bf J}^\pm_{\rm pola} \simeq \frac{m_e\gamma_\pm S}{|{\bf E}|}\hat{\bf E},\quad \rho^\pm_{\rm pola}\simeq \pm \frac{m_e\gamma_\pm |{\bf v}_\pm| S}{|{\bf E}|},
\label{pcurrent}
\end{align}
where the magnetic field ${\bf B}=0$. In Eq.~(\ref{me}), $J_{\rm ext}^{\mu} = (\rho_{\rm ext},{\bf J}_{\rm ext})$ is an external electric current.
\noindent
{\it Basic equations of motion.}
\hskip0.1cm
For simplicity to start with, we consider the electric field ${\bf
E}_{\rm ext}$ created by a capacitor made of two parallel plates,
one carries an external charge $+Q$ and another $-Q$.
The sizes of two parallel plats are $L_x$ and
$L_y$, which are much larger than their separation $\ell$ in the
$\hat {\bf z}$-direction, i.e., $L_x\gg \ell$ and $L_y\gg \ell$. For
$|z|\sim {\mathcal O}(\ell)$, the system has an approximate
translation symmetry in the $(x,y)$ plane.
As results the electric field ${\bf E}_{\rm ext}(x,y,z)\approx E_{\rm ext}(z)\hat{\bf z}$ and ${\bf B}_{\rm ext}(x,y,z)\approx 0$, is approximately homogeneous in the $(x,y)$ plane and confined within the capacitor. 
In addition, $\partial {\bf E}_{\rm ext}/\partial t\approx 0$, namely, this electric field is assumed to be continuously supplied by an external source $(+Q,-Q)$ or slowly varying.

In order to do calculations we model this electric field as the one-dimensional Sauter electric
field in the $\hat{\bf z}$-direction
\begin{equation}
E_{\rm ext}(z)=E_0/\cosh^2(z/\ell), \quad
\sigma \equiv e E_0 \ell/m_e c^2=(\ell/\lambda_C)(E_0/E_c),
\label{sfield}
\end{equation}
where the $\lambda_C$ is Compton wavelength, the external electric charge is given by $\partial E_{\rm ext}(z)/\partial z=4\pi\rho_{\rm ext}$ and the external electric current vanishes $J_{\rm ext}=0$ for the field being static $\partial E_{\rm ext}/\partial t=0$.
In the electric field configuration (\ref{sfield}) and ${\bf B}\approx 0$,
the ``averaged'' velocities $v_\pm$ of
electrons and positrons fluids are in the $\hat {\bf z}$-direction,%
\begin{equation}
U_\pm^{\mu}=\gamma_\pm\left(  1,0,0,\pm v_\pm\right),
\label{fourv}
\end{equation}
and the total fluid current- and charge-densities (\ref{me}) $J^\mu=(\rho,{\bf J})$ are
\begin{align}
J_z &= en_{+}v_{+}+en_{-}v_{-}
+\frac{m_e(\gamma_++\gamma_-) S}{E},\label{tcur}\\
\rho &= e\left(n_{+}-n_{-}\right)+\frac{m_e(\gamma_+v_+-\gamma_-v_-) S}{E}.
\label{cac}
\end{align}
The system can be approximately treated as a $1+1$ dimensional system in terms of space-time variables $(z,t)$, and Eqs.~(\ref{contp}-\ref{me}) become for zero pressure \cite{pressure},
\begin{align}
\frac{\partial n_\pm}{\partial t}&\pm \frac{\partial n_\pm
v_\pm}{\partial z}=S,\label{nudot}\\
\frac{\partial\epsilon_\pm}{\partial t}&\pm
\frac{\partial p_\pm}{\partial z}= en_\pm v_\pm E+m_e\gamma_\pm S,\label{rhoudot}\\
\frac{\partial p_\pm}{\partial t}&\pm \frac{\partial p_\pm
v_\pm}{\partial z}=en_\pm E + m_e\gamma_\pm v_\pm S,\label{pudot}\\
\frac{\partial E}{\partial t}&=-4\pi J_z 
,\label{Eudot}\\
\frac{\partial E}{\partial z}&=4\pi (\rho
+\rho_{\rm ext}).\label{Ediv}
\end{align}
The total electric field $E(z,t)$ in Eqs.~(\ref{nudot}-\ref{Eudot}) is the superposition of two components:
\begin{align}
E(z,t)=E_{\rm ext}(z)+ E_{\rm ind}(z,t), \label{tote}
\end{align}
where the space- and time-dependent $E_{\rm ind}(z,t)$ is the electric field created by electron and positron pairs. We call $J_z(z,t)$ (\ref{tcur}), $\rho(z,t)$ (\ref{cac}) and $E_{\rm ind}(z,t)$ pair-induced electric current, charge and field.

As for the pair-production rate $S$ in Eqs.~(\ref{nudot}-\ref{Eudot}), instead of the
pair-production rate (\ref{srate}) for a constant field $E_0$, we adopt the following $z$-dependent formula for the pair-production rate in the Sauter
field (\ref{sfield}), obtained by using the WKB-method to calculate the probability of quantum-mechanical tunneling \cite{Hagen},
\begin{align}
S(z)
=\frac{m_e^4}{4\pi^3}\frac{E_0E(z)}{E^2_c\tilde{G}[0,\E]}e^{-\pi
G[0,\E]E_c/E_0},
\label{hrate}
\end{align}
where $G(0,\E)$ and $\tilde G(0,\E)$ are functions of the
energy-level crossings $\E(z)$ and we approximately adopt $E(z)\approx E_0/G(0,\E)\approx E_0/\tilde G(0,\E)$ in Eq.~(\ref{hrate}) in order to do feasible numerical calculations. As shown by the Fig.~2 in Ref.~\cite{Hagen}, the deviation of the pair-production rate (\ref{hrate}) due to this approximation is small.
\comment{\setlength\arraycolsep{0.5pt}
\begin{eqnarray}
G(0,\varepsilon)&=&2\sigma^2-\sigma
\sqrt{(\sigma-\E)^2-1}-\sigma
\sqrt{(\sigma+\E)^2-1},\\
\tilde{G}(0,\E)&=&\frac{\sigma}{2}\left(
\frac{1}{\sqrt{(\sigma-\E)^2-1}}+\frac{1}{
\sqrt{(\sigma+\E)^2-1}}\right),\nonumber
\end{eqnarray}
}
The formula (\ref{hrate}) is
derived for the static Sauter field (\ref{sfield}). However, analogously to the discussions for the plasma oscillations in spatially homogeneous fields \cite{5}-\cite{ion2}, it can be approximately
used for a time-varying electric field $E(z,t)$ (\ref{tote}),
provided the time-dependent component $E_{\rm ind}(z,t)$, created by
electron-positron pair-oscillations, varies much slowly compared with the rate of electron-positron pair-productions ${\mathcal O}(m_ec^2/\hbar)$.
This can be justified by
the inverse adiabaticity
parameter \cite{book1}-\cite{Popov},%
\begin{equation}
\eta=\frac{m_e}{\omega}\frac{E_0}{E_{c}}\gg1,
\label{eta}
\end{equation}
where $\omega$ is the frequency of pair-oscillations.

Eqs.~(\ref{nudot},\ref{rhoudot},\ref{pudot}) describe the motion
of electron-positron plasma coupling to the electric field $E$ and
source $S$ of pair-productions. The Maxwell equations
(\ref{Eudot},\ref{Ediv}) describe the motion of the electric field
(\ref{tote}) coupled to the current- and charge-densities
(\ref{cac}), leading to the wave equation of the propagating
electric field $E_{\rm ind}(z,t)$ \cite{Jackson},
\begin{align}
\frac{\partial^2 E_{\rm ind}}{\partial
t^2}-\frac{1}{c^2}\frac{\partial^2 E_{\rm ind}}{\partial z^2}=4\pi
\left(\frac{\partial \rho}{\partial z}+\frac{1}{c^2}\frac{\partial
J_z}{\partial t}\right),
\label{we1}
\end{align}
where we use $\partial E_{\rm ext}/\partial z =4\pi \rho_{\rm ext}$ and $\partial E_{\rm ext}/\partial t=0$. This wave equation shows the propagating electric field
$E_{\rm ind}(z,t)$ in the region ${\mathcal R}$ where the non-vanishing current $J_z$ and charge $\rho$ are, and both the propagation and polarization of the
electric field are in the $\hat {\bf z}$-direction. This implies a wave transportation
of electromagnetic energies inside the region ${\mathcal R}$. Since the current- and charge-densities ($\rho, J_z$) are functions of the field $E(t,z)$ (\ref{tote}), the wave equation is highly nonlinear, the dispersion relation of the field is very complex and the velocity of field-propagation is not the speed of light.
\noindent {\it Numerical integrations.} \hskip0.1cm Given the
parameters $E_0=E_c$ and $\ell =10^5\lambda_C$ of the Sauter field
(\ref{sfield}) as an initial electric field $E_{\rm ext}$,  we
numerically integrate Eqs.(\ref{nudot}-\ref{Eudot}) in the spatial
region ${\mathcal R}$: $-\ell/2\leq z \leq \ell/2$ and time interval
${\mathcal T}$: $0\leq t \leq 3500\tau_C$, where $\tau_C$ is the
Compton time. The value ${\mathcal T}\leq 3500\tau_C$  is chosen so
that the adiabatic condition (\ref{eta}) is satisfied, and the
spatial range ${\mathcal R}$  is determined by the capacity of
computer for numerical calculations. The electric field strength
$E_0$ is chosen around the critical value $E_c$, so that the
semiclassical pair-production rate (\ref{hrate}) can be
approximately used. Actually, $E_0, \ell$ and ${\mathcal T}$ are
attributed to the characteristics
of external 
ultra-strong electric fields $E_{\rm ext}$ established by either
experimental setups or astrophysical conditions.

\begin{figure}[!h]
\begin{center}
\includegraphics[width=4.5in]{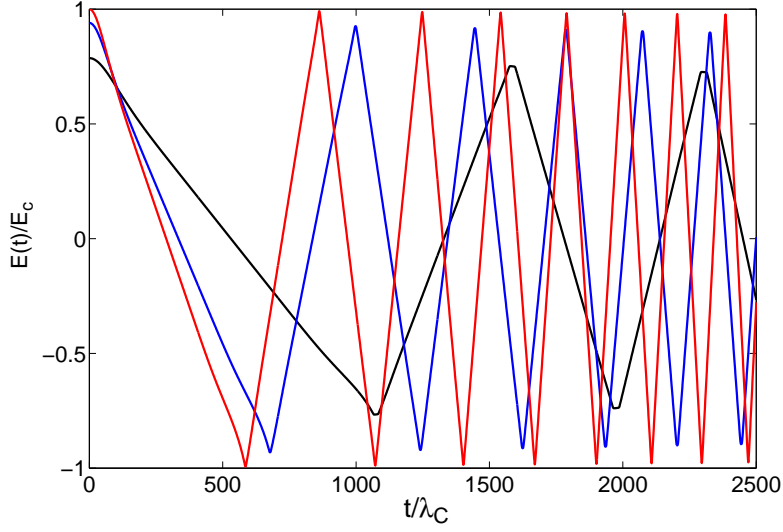}
\caption{Electric fields $E(z,t)$ are plotted as functions of $t$ at
three different points: $z=0$ (red), $z=\ell/4 $ (blue) and
$z=\ell/2$ (black). Analogously to the plasma oscillation in
homogeneous fields, the stronger initial field-strength, the larger
field-oscillation frequency is, i.e.,
$\omega(z=0)>\omega(z=\ell/4)>\omega(z=\ell/2)$, where $\omega(z)$
is the field oscillating frequency at the spatial point $z$.}
\label{Et}
\end{center}
\end{figure}

\begin{figure}[!h]
\begin{center}
\includegraphics[width=4.5in]{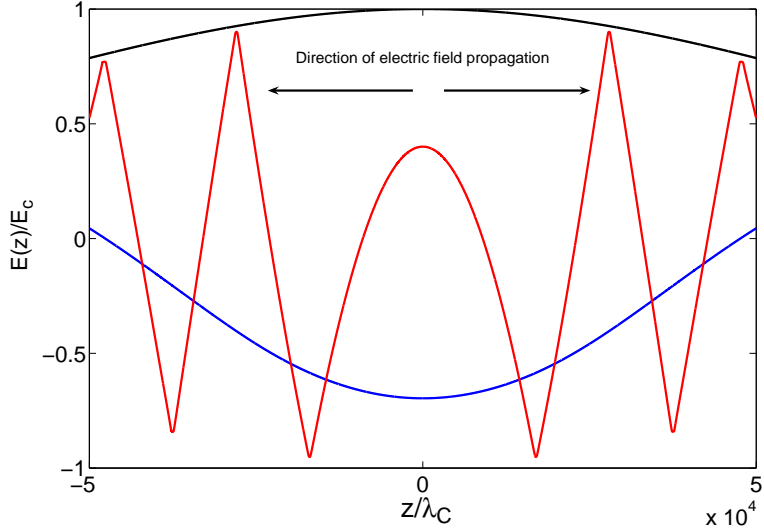}
\caption{Electric fields $E(z,t)$ are plotted as functions of $z$ at
three different times in the Compton unit: $t=1$(black), $t=500$
(blue) and $t=1500$ (red). As shown in Fig.~\ref{Et}, the electric
field $E(z,t)$ oscillation at the center ($z=0$) is most rapid, and
gets slower and slower at spatial points ($|z|>0$) further away from
the center. This implies the electric field wave propagating in the
space, and the directions of propagations are indicated. }
\label{Ex}
\end{center}
\end{figure}

\begin{figure}[!h]
\begin{center}
\includegraphics[width=4.5in]{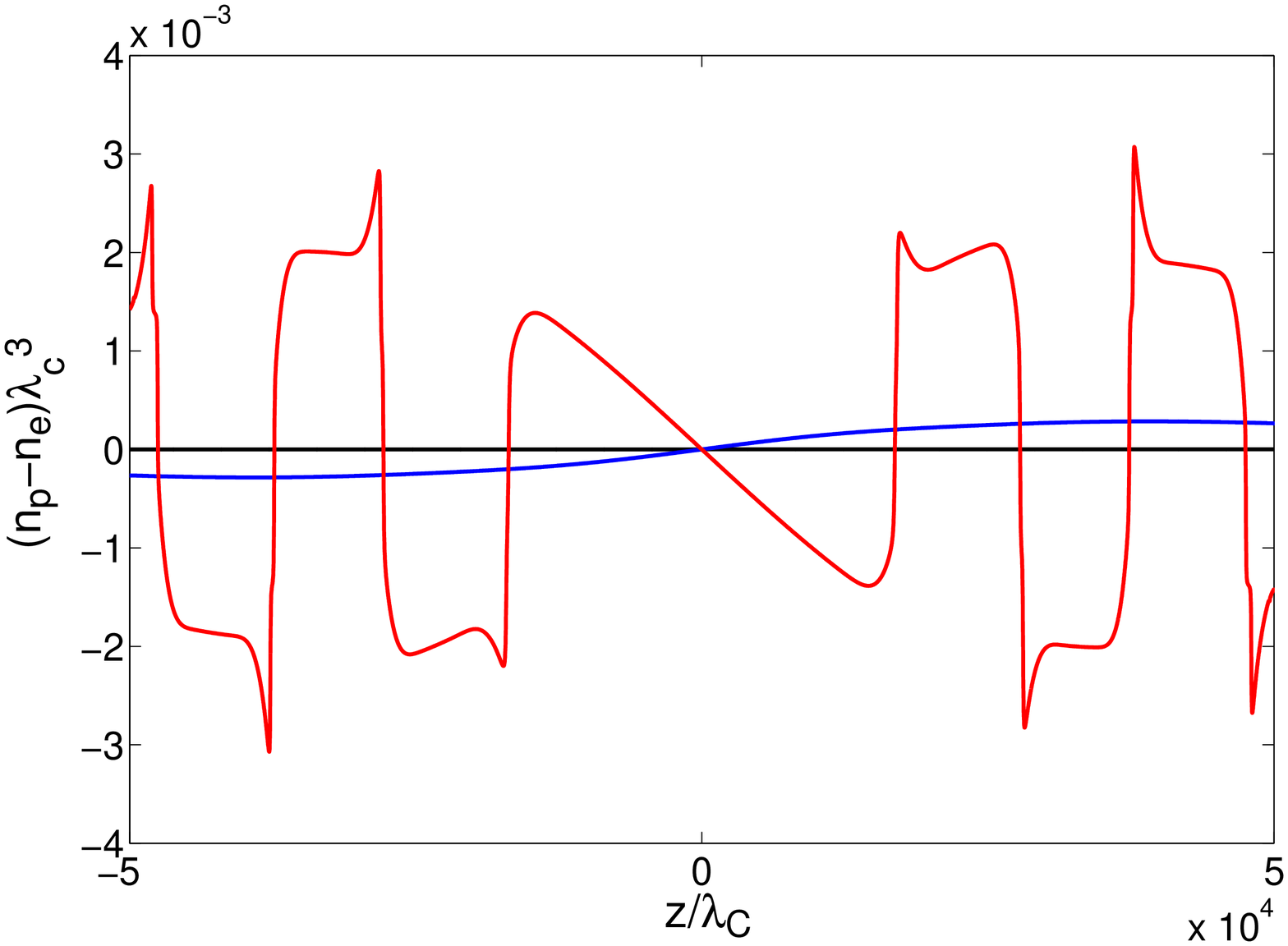}
\caption{The net charge density $\rho(z,t)$ [see Eq.~(\ref{cac})] as
a function of $z$ at three different times: $t=1$ (black, nearly
zero), $t=500$ (blue) and $t=1500$ (red). It is shown that the net
charged density value $|\rho(z,t)|$ is zero at the center where the
initial electric field gradient vanishes [see Eq.~(\ref{Ediv})],
whereas it increases as the initial electric field gradient
increases for $|z|>0$.
}\label{nx}
\end{center}
\end{figure}

\begin{figure}[!h]
\begin{center}
\includegraphics[width=4.5in]{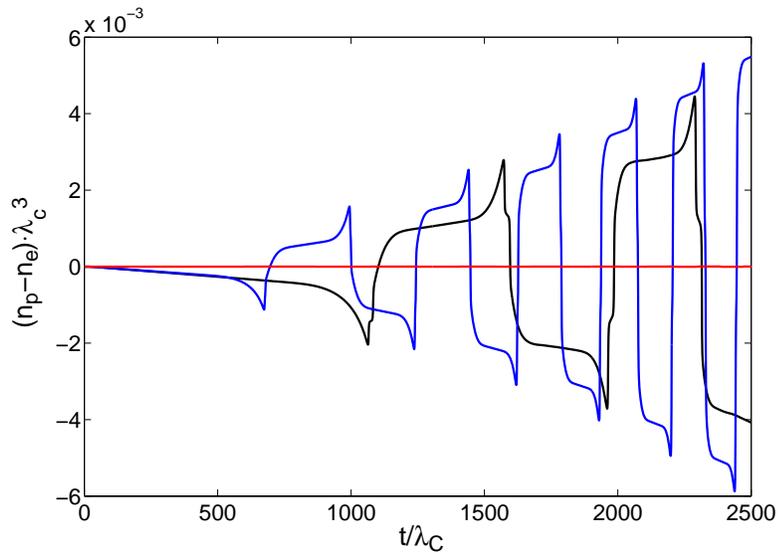}
\caption{The net electric charge density $\rho(z,t)$ [see Eq.~(\ref{cac})]
as a function of $t$ at three different points:
$z=0$ (red, nearly zero) , $z=\ell/4$ (blue) and $z=\ell/2$ (black).
It is shown that the net electric charge density $\rho(z,t)$
(except the center $z=0$) increases as time.
}\label{nt}
\end{center}
\end{figure}

\begin{figure}[!h]
\begin{center}
\includegraphics[width=4.5in]{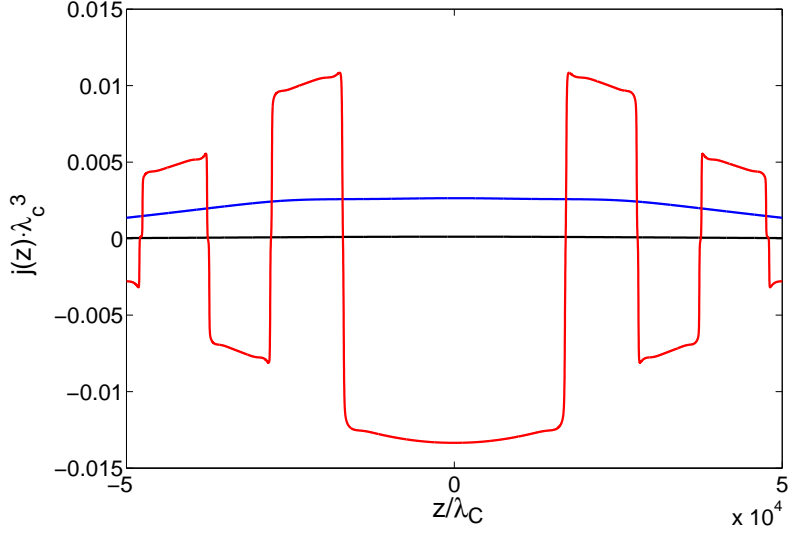}
\caption{Electric current densities $j_z(z,t)$ [see
Eq.~(\ref{tcur})] as functions of $z$ at three different times:
$t=1$ (black), $t=500$ (blue) and $t=1500$ (red). Following
Eq.~(\ref{Eudot}), the electric current alternates following the
alternating electric field (see Fig.~\ref{Et}), the plateaus
indicate the current saturation for $v\sim c$ and its spatial
distribution is determined by the initial electric field $E_{\rm
ext}(z)$. }\label{jx}
\end{center}
\end{figure}

\begin{figure}[!h]
\begin{center}
\includegraphics[width=4.5in]{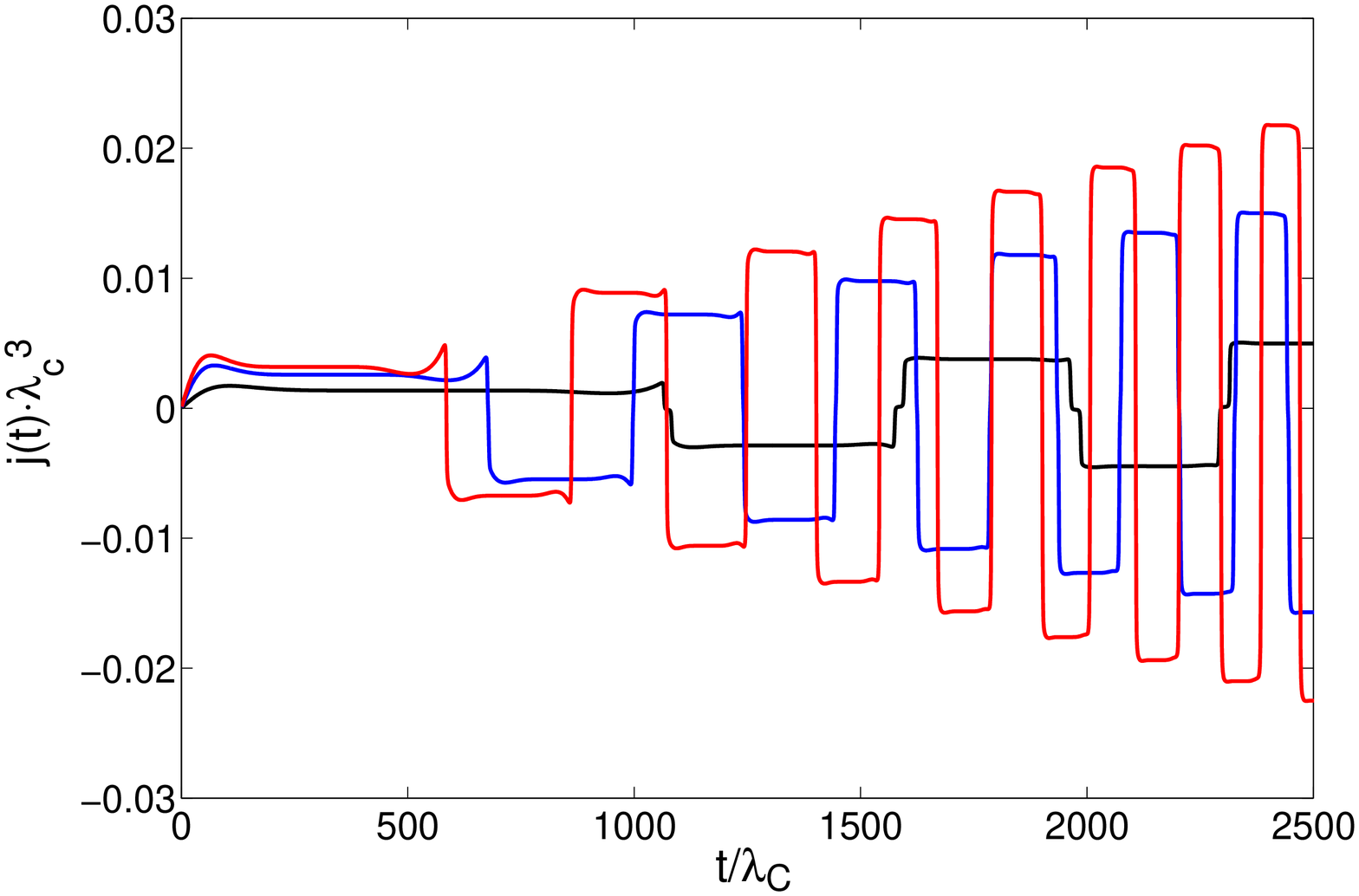}
\caption{Electric current densities $j_z(z,t)$ [see Eq.~(\ref{tcur})] as functions of $t$ at three different points: $z=\ell/2$ (black), $z=\ell/4$ (blue) and
$z=0$ (red). The plateaus (see also Fig.~\ref{jt}) for the current saturation values increases as time, mainly due to the number-densities $n_\pm$ of electron-positron pairs increase with time. In addition, they are maximal at the center $z=0$ where the initial electric field is maximal, and decrease as the initial electric field $E_{\rm ext}(z)$ decreasing for $|z|>0$.}
\label{jt}
\end{center}
\end{figure}

In Figs.~\ref{Et}
and \ref{Ex}, we respectively plot the time- and space-evolution of the
total electric fields $E(z,t)$ (\ref{tote}) as functions of $t$ and $z$ at three different spatial points and times. As discussed in Figure captions, numerical results show the properties of the electric field wave $E_{\rm ind}(z,t)$ propagating in the plasma of oscillating electron-positron pairs, as described by the wave equation (\ref{we1}). This electric field wave propagates along the directions in which external electric field-strength decreases.
The wave propagation is rather complex, depending on the space and time variations of the net charge density $\rho(z,t)$ and current density $j_z(z,t)$, as shown in Figs.~\ref{nt}-\ref{jx}.
The net charge density $\rho$ oscillates (see Figs.~\ref{nx}
and \ref{nt}) proportionally to the field-gradient (\ref{Ediv}) and
at the center $z=0$ the charge density and field-gradient are zero
independent of time evolution (see Fig.~\ref{nt}). However, the total charge
of pairs $Q=\int_{\mathcal R} d^3x \rho $ must be zero at any time,
as required by the neutrality.
The electric
current $j_z(z,t)$ alternating in space and time follows the space and time evolution of the electric field $E(z,t)$
see Eq.~(\ref{Eudot}), as shown in Figs.~\ref{jx} and \ref{jt}.

We recall the discussions of the plasma oscillations in the case of
spatially homogeneous electric field $E_0$
without boundary \cite{5,gregory}. Due to the spatial
homogeneity of electric fields and pair-production rate $S$
(\ref{srate}), the number-densities $n_\pm(t,{\bf x})=n(t)$
(\ref{meann}), ``averaged'' velocities $|{\bf v}_\pm(t,{\bf
x})|=v(t)$ (\ref{meanv}) and energy-momenta
$\epsilon_\pm(t,{\bf x})=\epsilon(t),|{\bf p}_\pm(t,{\bf x})|=p(t)$
(\ref{led}) are spatially homogeneous
so that the charge density (\ref{cac}) $\rho\equiv 0$ identically
vanishes and current (\ref{tcur}) $J_z=J_z(t)$. All spatial
derivative terms in Eqs.~(\ref{nudot}-\ref{pudot}) and
Eq.~(\ref{we1}) vanish and Eq.~(\ref{Ediv}) becomes irrelevant. As
results, the plasma oscillations described is the oscillations of electric fields and currents with respect time at
each spatial point, and the electric field has no any spatial
correlation and does not propagate.

In contrary to the plasma oscillation in homogeneous fields, the
presence of such field-propagation in inhomogeneous fields is due
to: (i) non-vanishing field-gradient $\partial_z E$ (\ref{Ediv}) and
net charge-density $\rho$ (\ref{cac}), as shown in Figs.~\ref{nx}
and \ref{nt}, give the spatial correlations of the fields at
neighboring points; (ii) the stronger field-strength, the larger
field-oscillation frequency is, as shown in Fig.~\ref{Et}; (iii) at
the center $z=0$ the field-strength is largest and the
field-oscillation is most rapid, and the field-oscillations at
points $|z|>0$ are slower and in retard phases, as shown in
Fig.~\ref{Ex}. The point (i) is essential, the charge density $\rho$
oscillates (see Figs.~\ref{nx} and \ref{nt}) proportionally to the
field-gradient Eq.~(\ref{Ediv}) and at the center $z=0$ the charge
density and field-gradient are zero independent of time evolution
(see Fig.~\ref{nt}). Such field-propagation is reminiscent of the
drift motion of particles driven by a field-gradient (
``ponderomotive'') force, which is a cycle-averaged force on a
charged particle in a spatially inhomogeneous oscillating
electromagnetic field \cite{ponderomotive}.

\comment{
By comparing evolution of the total electric field at different
points, we can find the frequency of field oscillation depends on
the strength of the external field: the point with stronger external
electric field has faster oscillation frequency. This is easy to
understand, the stronger external field will product more pairs,
accelerate them more fast. Then the pairs oscillate more fast. (See
Fig.\ref{Et})
}
\comment{
In Fig.\ref{netcharge}, we have explained that due to the spatially
inhomogeneous Sauter-style electric field, there are net charges at
every local slice. The numerical results prove our hypothesis. And
also, as time growing, the difference of number densities of
positrons and electrons becomes bigger and bigger. So, this is why
the total electric field distorts Sauter-style more and more.
In Fig.\ref{nx} we show that the difference of number densities of
positron and electron through three different time. Attending that
the total net charge is zero during whole evolution.
}

\comment{
In Fig.\ref{nt}, the evolution of number density difference of pairs
at three different points is shown. We can find that the number
densities of positron and electron are almost same at the central
point of electric field in whole evolution time. But far away from
the center, the difference of $n_+$ and $n_-$ becomes bigger and
bigger.
In conclusion, from Fig.\ref{nx} and Fig.\ref{nt}, it is very clear
that the difference of number densities of $e^+$ and $e^-$
distributes in spatial coordinate and evolves in time too. More far
away from the central point, more long evolution time, the net
charges become much more.
}
\comment{
\subsection{Evolution of current density}
The net charger-number density at local area is usually much smaller
than the number density of positron or electron, because from
Fig.\ref{nx},\ref{nt}, $n_+\sim n_-$, and the net charger density
should be $q=e(n_+-n_-)$. But for electronic current, since the
direction of positron and electron movement is opposite, so the
total current density induced by pairs movement should be
$j_{cond}=e(v_+n_++v_-n_-)$ and almost is twice of the current only
counting positrons or electrons. Also, the total net charge in the
whole area is zero, but usually the total current integrated in the
whole area is not zero. So the variation of current will product a
radiation field at a far point.
In Fig.\ref{jt} we show the current distribution in space at three
different time: $t=1,48$ and $120$.
}
\comment{
From Fig.\ref{jt}, it is obviously different with constant field. In
constant electric field, at one moment, the strength and direction
of current is uniform at everywhere \cite{4},\cite{gregory}, only
oscillating with time. But for spatially inhomogeneous Sater-style
field, the current density not only oscillate in time but also is a
function of position. Furthermore, as time growing, the whole field
area is divided into many smaller areas, and the direction of
current in these smaller areas is changed one by one.
In Fig.\ref{jx}, it presents the evolution of current densities
along time at three different point: $x=0$, $x=l/4$ and $x=l/2$. We
can find for fixed position, the current evolution in time is
similar with the results of \cite{4}.
}

\begin{figure}[!h]
\begin{center}
\includegraphics[width=4.5in]{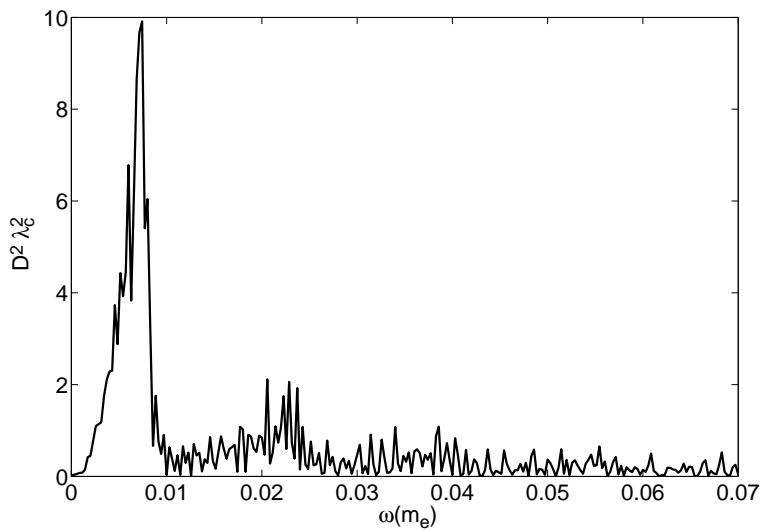}
\caption{In the Compton unit, normalizing $\tilde D(\omega)$ [see
Eq.~(\ref{ftd})] by the volume ${\mathcal V}\equiv \int d^3{\bf x}'$
of the radiation source ${\bf J}(t',{\bf x}')$, we plot $|\tilde
D(\omega)|^2$ [see Eq.~(\ref{inten})] representing the narrow
energy-spectrum of the radiation field ${\bf E}_{\rm rad}$ and peak
locates at the frequency $\omega_{\rm peak}\approx 0.08m_e$.}
\label{sp}
\end{center}
\end{figure}

\begin{figure}[!h]
\begin{center}
\includegraphics[width=4.5in]{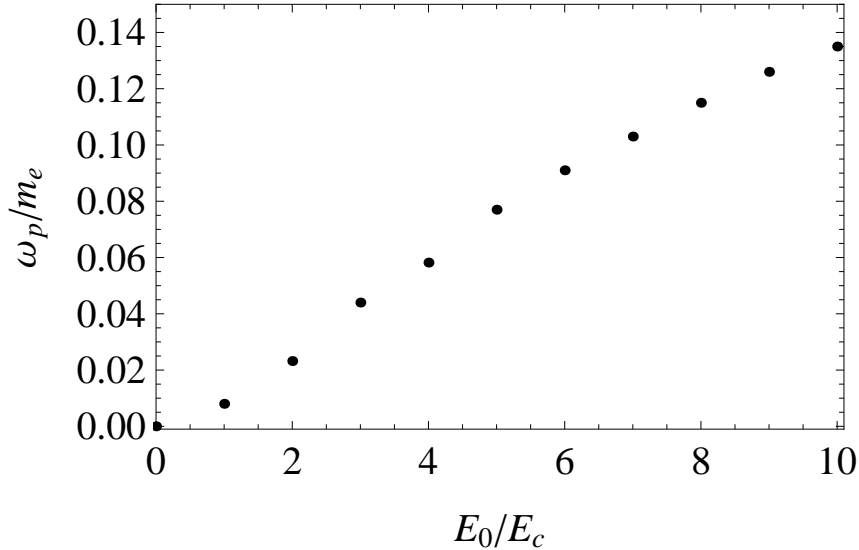}
\caption{The peak frequency $\omega_{\rm peak}$ of the radiation
approximately varies from $4$KeV to $70$ KeV as the initial electric
field strength $E_0$ varies from $E_c$ to $10E_c$. The values for
very large field-strengths $E_0/E_c >1$ possibly receive
corrections, since the semiclassical pair-production rate
(\ref{hrate}) is approximately adopted and the pressure term (see
\cite{pressure}) is not properly taken into account.} \label{S_E}
\end{center}
\end{figure}

\noindent
{\it Radiation fields.}
\hskip0.1cm
As numerically shown in Fig.~\ref{Et}-\ref{jt}, the propagation of the electric field wave $E_{\rm ind}(z,t)$ inside the region ${\mathcal R}$ is rather complex, due to th high non-linearity of wave equation (\ref{we1}). Nevertheless,
the electromagnetic radiation fields ${\bf E}_{\rm rad}$ and ${\bf B}_{\rm rad}$ far away from the
region ${\mathcal R}$ are completely determined and could be experimentally observable.
At the space-time point ($t, {\bf x}$) of an observer, the electromagnetic radiation fields ${\bf E}_{\rm rad}(z,t)$ and ${\bf B}_{\rm rad}(z,t)$,
emitted by the variations of electric charge density $\rho({\bf x}', t')$ and current-density ${\bf J}({\bf x}', t')$ in the region ${\mathcal R}$ $({\bf x}'\in {\mathcal R})$ and time $t'$ $(t'\in {\mathcal T})$, are given by \cite{Jackson}
\comment{
\begin{eqnarray}
{\bf E}_{\rm rad}(t,{\bf x})&=&-\int_{\mathcal R} d^3{\bf x}'
\frac{1}{R}\left[\nabla'\rho(t',{\bf
x}')+\frac{1}{c^2}\frac{\partial {\bf J}(t',{\bf x}')}{\partial
t'}\right]_{\rm ret},
\label{retare}\\
{\bf B}_{\rm rad}(t,{\bf x})&=&\int_{\mathcal R} d^3{\bf x}'
\frac{1}{cR}\left[\nabla'\times {\bf J}(t',{\bf x}')\right]_{\rm
ret}, \label{retarb}
\end{eqnarray}
}
\begin{align}
{\bf E}_{\rm rad}(t,{\bf x})= &-\int_{\mathcal R} d^3{\bf x}' \Big\{\frac{\hat{\bf R}}{R^2}\left[\rho(t',{\bf x}')\right]_{\rm ret}+\frac{\hat{\bf R}}{cR}\left[\frac{\partial \rho(t',{\bf x}')}{\partial t'}\right]_{\rm ret}\nonumber\\
&+\frac{1}{c^2R}\left[\frac{\partial {\bf J}(t',{\bf x}')}{\partial t'}\right]_{\rm ret}\Big\},\label{retare1}\\
{\bf B}_{\rm rad}(t,{\bf x})= &\int_{\mathcal R} d^3{\bf x}' \left\{\left[{\bf
J}(t',{\bf x}')\right]_{\rm ret}\times \frac{\hat{\bf
R}}{cR^2}+\left[\frac{\partial {\bf J}(t',{\bf x}')}{\partial
t'}\right]_{\rm ret}\times\frac{\hat{\bf R}}{c^2R}\right\}.
\label{retarb1}
\end{align}
where the subscript ``${\rm ret}$'' indicates $t'=t-R/c$, $R=|{\bf
x}-{\bf x}'|$.
\comment{ In the near zone, we select a point ${\bf x}=(0,0,z)$ and
$|z|\ll {\rm min}(L_x, L_y)$  for discussions.
The radiative magnetic field approximately vanishes ${\bf B}_{\rm
rad}(t,z)\approx 0$, since the sum over all contributions from the
region ${\mathcal R}$ cancel each other, due to the approximate
translation symmetry in the ($x,y$) plane. The electric field is
propagating and polarized in the $\hat{\bf z}$-direction
\begin{align}
E_{\rm rad}(t,z)&=\int d^3{\bf x}' \frac{1}{R}\left(\frac{\partial \rho}{\partial z'}+\frac{1}{c^2}\frac{\partial J_z}{\partial t'}\right)_{t'=t-R/c},\nonumber\\
&\approx L_xL_y\ell \frac{1}{z}\left(\frac{\partial \rho(t',{\bf
x}')}{\partial z'}+\frac{1}{c^2}\frac{\partial J_z(t',{\bf
x}')}{\partial t'}\right)_{t'=t-z/c;\hskip0.1cm {\bf x}'=0},
\label{retare1}
\end{align}
where we assume the homogeneous variations of $\rho(t',{\bf x}')$
and $J_z(t',{\bf x}'))$ in the region ${\mathcal R}$. we should
be able to numerically calculate and plot this ... }
In the radiation zone $|{\bf x}|\gg |{\bf x}'|$ and $R\approx|{\bf x}|$, where is far away from the plasma oscillation region ${\mathcal R}$
, the radiation fields (\ref{retare1},\ref{retarb1}) approximately are
\begin{align}
{\bf E}_{\rm rad}(t,{\bf x})&\approx-\frac{1}{c^2|{\bf x}|}\int
d^3{\bf x}' \left[\frac{\partial {\bf J}(t',{\bf x}')}{\partial
t'}\right]_{\rm ret},
\label{retaref}\\
{\bf B}_{\rm rad}(t,{\bf x})&\approx \hat{\bf R}\times{\bf E}_{\rm
rad}(t,{\bf x}), \label{retarbf}
\end{align}
where we use the
charge conservation (\ref{current}) and total
neutrality condition of pairs $\int_{\mathcal R} d^3{\bf x}'
\rho(t',{\bf x}')=0$.
The first terms in Eqs.~(\ref{retare1},\ref{retarb1}) are the Coulomb-type
fields decaying away as ${\mathcal O}(1/|{\bf x}|^2)$.
The Fourier transforms of
Eqs.~(\ref{retaref}) and (\ref{retarbf}) are
\begin{align}
\tilde {\bf E}_{\rm rad}(\omega,{\bf x})&\approx -\frac{e^{-ik |{\bf x}| }}{c^2|{\bf x}|}\tilde {\bf D}(\omega),\quad \tilde {\bf B}_{\rm rad}(\omega,{\bf x})\approx  \hat{\bf R}\times\tilde{\bf E}_{\rm rad}(\omega,{\bf x})\label{fte1}\\
\tilde {\bf D}(\omega) &\equiv \int_{\mathcal
R} d^3{\bf x}' \int_{\mathcal T} dt' e^{i\omega t'}\left[\frac{\partial {\bf
J}(t',{\bf x}')}{\partial t'}\right],
\label{ftd}
\end{align}
where the wave number $k=\omega/c$ and the numerical integration (\ref{ftd})
is carried out overall the space-time evolution of
the electric current ${\bf J}({\bf x}',t')$ (see Figs.~\ref{jt} and \ref{jx}).
For definiteness we thinks of the oscillation currents occurring for some finite interval of time ${\mathcal T}$ or at least falling off for remote past and future times, so that the total energy radiated is finite, thus the energy radiated per unit solid angle per frequency interval is given by \cite{Jackson}
\begin{equation}
\frac{d^2I}{d\omega d\Omega}=2|\tilde {\bf D}(\omega)|^2.
\label{inten}
\end{equation}
The squared amplitude $|\tilde {\bf D}(\omega)|^2$ as a function of
$\omega$ gives the spectrum of the radiation (see Fig.~\ref{sp}),
which is very narrow as expected with a peak locating at
$\omega_{\rm peak}\approx 0.08m_e= 4$KeV for $E_0=E_c$, consistently
with the plasma oscillation frequency (see Fig.~\ref{Et}). The
energy-spectrum and its peak are shifted to high-energies as the
initial electric field-strength increases, and the relation between
the spectrum peak location and the electric field-strength is shown
in Fig.~\ref{S_E}. In addition, the energy-spectrum and its peak are
also shifted to high-energies as the temporary duration ${\mathcal
T}$ of plasma oscillations increases (see Fig.~\ref{Et}).
In calculations, the temporary duration ${\mathcal T}=3500\tau_C$ is
chosen, not only to satisfy the adiabaticity condition
Eq.~(\ref{eta}) \cite{ceta}, but also to be in the time duration
when the oscillatory behavior is distinctive (see
Figs.~\ref{Et},\ref{nt},\ref{jt}), since the oscillations of
pair-induced currents damp and pairs annihilate into photons
\cite{5}. The radiation intensity (\ref{inten}) depends on the
strength, spatial dimension and temporal duration of strong external
electric fields, created by either experimental setups or
astrophysical conditions.

\comment{
The electromagnetic Poynting vector and radiation energy-densities are,
\begin{eqnarray}
{\bf S}=\frac{c}{4\pi}{\bf E}_{\rm rad}\times {\bf B}_{\rm
rad}=c\E_{\rm em}\hat{\bf R},\quad \E_{\rm em}=\frac{1}{8\pi}({\bf
E}_{\rm rad}^2+{\bf B}_{\rm rad}^2)=\frac{1}{4\pi}{\bf E}_{\rm
rad}^2, \label{se}
\end{eqnarray}
which crucially depends on the spatial region ${\mathcal R}$ and time interval ${\mathcal T}$ of the plasma oscillations that we do not discuss in this paper.
The
total radiation energy flux of radiation $F=4\pi R^2|{\bf S}|=c|{\bf
x}|^2|{\bf E}_{\rm rad}|^2$.
and total radiation energy
\begin{eqnarray}
\E_{\rm tot}=\int dt' c|{\bf x}|^2|{\bf E}_{\rm rad}|^2= \int
dt'\left|\int d^3{\bf x}' \left[\frac{\partial {\bf J}(t',{\bf
x}')}{\partial t'}\right]\right|^2, \label{tot}
\end{eqnarray}
where $\int dt'$ is the integration over the period of plasma
oscillations, which depends on the timescale of external sources
that make large electric fields.
}

\vskip0.1cm \noindent {\it Conclusions and remarks.} \hskip0.1cm
We show the space and time evolutions of pair-induced
electric charges, currents and fields in strong external electric
fields bounded within a spatial region. These results imply the wave
propagation of the pair-induced electric field and
wave-transportation of the electromagnetic energy in the strong
field region. Analogously to the electromagnetic radiation emitted
from an alternating electric current, the space and time variations
of pair-induced electric currents and charges emit an
electromagnetic radiation. We show that this radiation has a
the peculiar energy-spectrum
(see Fig.~\ref{sp}) that is clearly distinguishable from the energy-spectra
of the bremsstrahlung radiation, electron-positron annihilation and
other possible background events.
This possibly
provides a distinctive way
to detect the radiative signatures for the production and
oscillation of electron-positron pairs in ultra-strong electric
fields that can be realized in either ground laboratories or
astrophysical environments.

As mentioned in introduction, the critical electric field $E_c$ will be reached soon in ground laboratories and sensible methods to detect signatures of pair-productions become important.
Recently, the momentum signatures of pair-production is found \cite{Dunne}
in a time-varying electric field $E(t)$ with sub-cycle structure. On the other hand, space-based telescopes the Swift-BAT \cite{Swift}, NuSTAR \cite{Nustar} and Astro-H \cite{AstroH} focusing high-energy X-ray missions, will also give possibilities of detecting
X-ray radiation signature, discussed in this paper, from compact stars with electromagnetic structure.

{\it Acknowledgements:} We thank H.~Keinert for helpful discussions
on the wave equation~(\ref{we1}).



\begin{thebibliography}{99}
\bibitem{pressure} For an electric field $E\sim E_c$, the number-density of electron-positron pairs is small and the pressure of pairs can be neglected. While for an over electric field $E\gg E_c$, the number-density of pairs is large and the collisions and annihilation of pairs into photons are important, leading to the energy equipartition of electron, positrons and photons. In this case, the pressure, effective temperature and equation of state have to be considered.
\bibitem{ceta}
We check the
two cases $E_0=E_{c}$ and $E_0=10E_{c}$, and find for the first
oscillation $\eta=865$ and $\eta=487$ respectively. As can be seen
for the Fig. \ref{Et} the frequencies $\omega$ of pair-oscillations
increase with time which means the parameter $\eta$ becoming
smaller. Eventually it may reach unity so the formula (\ref{hrate})
becomes inapplicable.
\end{thebibliography}
\end{document}